\documentclass[12pt]{iopart}
\usepackage{graphicx}
\begin{document}

\title[Anomalous relaxation in binary mixtures: a dynamic facilitation picture]
{Anomalous relaxation in binary mixtures: \\ a dynamic facilitation picture}

\author{A J Moreno$^1$ and J Colmenero$^{1,2,3}$}
\address{$^1$ Donostia International Physics Center, Paseo Manuel de Lardizabal 4,
20018 San Sebasti\'{a}n, Spain.}
\address{$^2$ Departamento de F\'{\i}sica de Materiales, Universidad del Pa\'{\i}s Vasco (UPV/EHU),
Apartado 1072, 20080 San Sebasti\'{a}n, Spain.}
\address{$^3$ Unidad de F\'{\i}sica de Materiales, Centro Mixto CSIC-UPV, 
Apartado 1072, 20080 San Sebasti\'{a}n, Spain.}
\ead{wabmosea@sq.ehu}

\begin{abstract}
Recent computational investigations in polymeric and non-polymeric binary mixtures
have reported anomalous relaxation features when both components exhibit very different
mobilities. Anomalous relaxation is characterized by sublinear power law behaviour 
for mean squared displacements, logarithmic decay in dynamic correlators,
and a striking concave-to-convex crossover in the latters by tuning the relevant control parameter,
in analogy with predictions of the Mode Coupling Theory for state points close to higher-order transitions.
We present Monte Carlo simulations on a coarse-grained model for relaxation in binary mixtures. 
The liquid structure is substituted by a three-dimensional array of cells. A spin variable
is assigned to each cell, representing unexcited and excited local states of a mobility field. 
Changes in local mobility (spin flip) are permitted according 
to kinetic constraints determined by the mobilities of the neighbouring cells. 
We introduce two types of cells (``fast'' and ``slow'') with very different rates 
for spin flip. This coarse-grained model qualitatively reproduces the mentioned anomalous relaxation
features observed for real binary mixtures.

\end{abstract}


\section{Introduction}

Binary mixtures of small and large particles with sufficiently large size disparity,
at low or moderate concentrations of the small particles,
exhibit a large separation in the time scales for both components.
Recent computational investigations in polymeric and non-polymeric mixtures of soft spheres
\cite{blendpaper,mixturepaper,disparpaper} in such conditions have reported
anomalous relaxation features at intermediate intervals of more than three time decades 
between the short-time ballistic and long-time diffusive regimes. 
These anomalous features are: i) sublinear power law behaviour for mean squared displacements,
with decreasing exponent by decreasing temperature, ii) logarithmic relaxation in dynamic correlators, 
iii) concave-to-convex crossover in dynamic correlators by tuning the relevant control parameter.
These results exhibit striking analogies with predictions of the Mode Coupling Theory (MCT)
for state points close to higher-order MCT transitions \cite{schematic1,schematic2,sperl},
which arise in regions of the control parameter space where competition between
different mechanisms for dynamic arrest takes place. For the fast component in binary mixtures
we have suggested a competition between confinement and bulk-like caging 
\cite{blendpaper,mixturepaper,disparpaper}, respectively induced by the host matrix formed
by the slow component, and by the surrounding particles belonging to the fast component.
This view is supported by the observed non-trivial analogies \cite{disparpaper} 
with a recent MCT theory for fluids confined in matrixes with interconnected voids 
\cite{krakoviack1,krakoviack2}, for which a higher-order point has been explicitly derived.
Striking analogies with anomalous relaxation features in short-ranged attractive colloids are also observed.
For these latter systems a higher-order MCT transition has been derived as the 
result of a competition between steric repulsion and reversible bond formation 
for dynamic arrest \cite{sperl,fabbian,bergenholtz,dawson}.

Kinetically constrained models \cite{fredrickson,sollich} are used as coarse-grained
pictures for relaxation in supercooled liquids. In these models, the liquid structure is substituted
by an array of cells. The cell size roughly corresponds
to a density correlation length. A spin variable is assigned to each cell, with values 0 or 1 denoting
respectively unexcited and excited local states in a mobility field. Changes in local mobility
(spin flips) for a given cell are permitted according to kinetic constraints 
determined by the mobilities of the neighbouring cells. Propagation of mobility, which yields
structural relaxation, occurs via dynamic facilitation: microscopic regions become temporarily
mobile only if neighbouring regions are mobile. Kinetically constrained models provide in a simple way
an important feature exhibited by glass-forming liquids: the growing of dynamic correlation lengths 
by decreasing temperature, leading to dynamic heterogeneity \cite{prnat,prlfac,prefac,nef}.

Some of the anomalous dynamic features exhibited by short-ranged attractive colloids,
and associated with a higher-order point within the framework of MCT, as logarithmic relaxation
or reentrant behaviour of the diffusivity \cite{sperl}, have been recently reproduced by a simple model 
based on dynamic facilitation \cite{geissler}. Motivated by this fact and by the mentioned dynamic 
analogies with binary mixtures, we investigate a simple kinetically constrained model aimed 
to reproduce qualitative relaxation features for such mixtures. 
In Section 2 we introduce the model and give computational details.
Simulation results are presented and discussed in Section 3. 

\section{Model and Computational Details}

We have carried out Monte Carlo (MC) simulations on a variation \cite{dynfacpaper} (see below) 
of the north-or-east-or-front (NEF) model recently investigated by Berthier and Garrahan \cite{nef}. 
In these three-dimensional model directionality in the kinetic constraints is imposed
to mimic fragile liquid behaviour \cite{nef}. Spin flip of a given cell is only permitted if there is 
at least one excited neighbouring cell in the north-or-east-or-front direction. 
More specifically, if the cell is denoted by its position vector ${\bf i}=$ ($i_{\rm x}$,$i_{\rm y}$,$i_{\rm z}$),
spin flip is permitted according to the following rules:

i) At least one of the neighbouring cells in the north ($i_{\rm x}$,$i_{\rm y}$,$i_{\rm z}+1$),
or east ($i_{\rm x}$,$i_{\rm y}+1$,$i_{\rm z}$),
or front ($i_{\rm x}+1$,$i_{\rm y}$,$i_{\rm z}$) direction is excited, i.e., it has spin 1.

ii) If i) is fulfilled, spin flip is accepted according to the Metropolis rule \cite{frenkel}.
Hence, $1 \rightarrow 0$ is always accepted, while $0 \rightarrow 1$ is accepted with a 
thermal probability $[1+\exp(1/T)]^{-1}$, where $T$ is the temperature. 

In the present work we investigate a binary mixture of cells with the same population
of excitations, $[1+\exp(1/T)]^{-1}$, but different rates for excitation,
$\exp(-E/T)[1+\exp(1/T)]^{-1}$, and decay, $\exp(-E/T)$, of the cell mobility.
This choice fulfills detailed balance. We use the activation energies $E =$ 0 and 3 for the different components, 
which are respectively denoted as ``fast'' and ``slow'' cells. Hence, spin flip rules for 
the fast component are the same as in the original NEF model, aimed to reproduce bulk-like dynamics.
The probability of spin flip for the slow cells is decreased by a factor $\exp(-3/T)$ as compared
to that of the fast cells, providing a large time scale separation for relaxation of both components (see below).  

The mixture composition, $x_{\rm f}$, is defined as the fraction of fast cells.
We investigate a wide range of composition and temperature as control parameters.
A square box of side $N = 50$ cells is used. Periodic boundary conditions
are implemented. Slow and fast cells are randomly distributed according to the selected value of $x_{\rm f}$.
Time is given in MC steps. Within each MC step a total of $N^3$ trials (one trial for each cell) is performed.
The configuration of the mobility field at times $t$ and $t+1$ is defined as that respectively before
and after the $N^3$ trials.

\section{Results and Discussion}

We have evaluated the mean squared displacement according to the definition 
given in Ref. \cite{yung}. A ``probe molecule'' placed in a given cell ${\bf i}$ at time $t$ is permitted 
to perform a jump ${\bf \Delta} = (\Delta_x, \Delta_y, \Delta_z)$, with $\Delta_{\alpha} \in \{-1,0,1\}$, 
to an adjacent cell ${\bf j} = {\bf i}+{\bf \Delta}$ provided that {\it both} ${\bf i}$ and ${\bf j}$
are instantaneously (i.e., at the same time $t$) in an excited state. 
Otherwise the jump is not permitted. If in the former case
there are several candidates between the adjacent cells, one of them is randomly selected.
The mean squared displacement is computed from an average over the trajectories of different (non-interacting)
probes initially placed in different cells. In the case of the binary mixture investigated in this work,
only jumps between cells of a same component (``fast'' $\rightarrow$ ``fast'' or ``slow'' $\rightarrow$ ``slow'')
are permitted. In this way the motion of a probe initially placed in a cell of a given component
remains coupled to the mobility field of that component.
Fig. 1 shows, at a fixed composition $x_{\rm f} = 0.33$, results for the mean squared displacement 
computed according to this procedure. The short-time ballistic regime exhibited by real systems 
is obviously absent due to the coarse-grained character of the model. 
A large time scale separation is induced between both components
at low temperatures. In analogy with the behaviour displayed by binary mixtures with
large size disparity \cite{blendpaper,mixturepaper,disparpaper}, an intermediate sublinear power 
law behaviour is observed for the fast component, and the corresponding exponent 
decreases by decreasing temperature.
The latter takes a value of about 0.6 at the lowest investigated temperature.

\begin{figure}
\begin{center}
\includegraphics[width=0.54\linewidth]{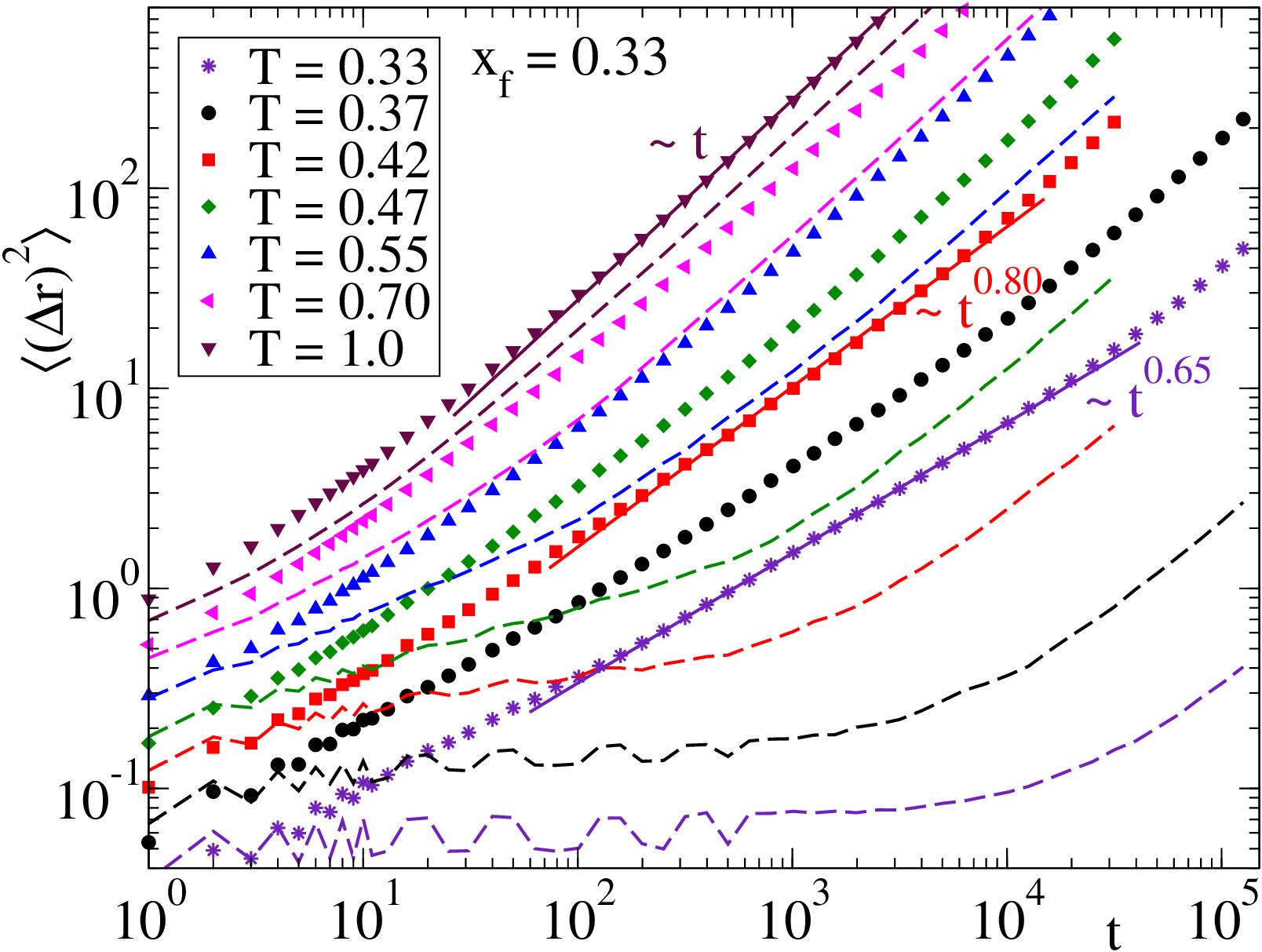}
\caption{Temperature dependence of the mean squared displacement at composition
$x_{\rm f}=0.33$. Symbols correspond to fast cells. Dashed lines correspond to slow cells
(same temperatures as for the former ones, decreasing from top to bottom).
Solid straight lines indicate linear or sublinear power law behaviour (exponents are given).}
\end{center}
\label{fig1}
\end{figure}

We have also computed the mean persistence function $P(t) = \sum_{{\bf i}} p({\bf i} ; t)/N^3$, 
where  $p({\bf i} ; t)$ is the persistence function of the cell ${\bf i}$ at time $t$.
The latter takes the value 1 if no spin flip has occurred for that cell in the interval [0,$t$]. 
If one or more spin flips have occurred,
it takes the value 0. According to the picture of dynamic facilitation, the decay of $P(t)$
is a signature of structural relaxation \cite{nef}.
Fig. 2 shows the temperature dependence of $P(t)$ at composition $x_{\rm f} = 0.33$.
The first decay usually observed for any correlator in real systems 
---corresponding to the transient regime at microscopic times--- is again absent due to coarse-graining.
As for the mean squared displacement, the introduction of very different rates for cell spin flip
yields very different relaxation times for fast and slow cells. 
While slow cells display standard relaxation, fast cells exhibit rather different relaxation features. 
In full analogy with results for dynamic correlators of the fast component
in binary mixtures \cite{blendpaper,mixturepaper,disparpaper}, $P(t)$ shows a concave-to-convex crossover
by decreasing temperature. At an intermediate temperature, pure logarithmic relaxation is observed over
a time interval of near three decades.
This behaviour is also observed by fixing temperature and varying the mixture composition (see Fig. 3).

\begin{figure}
\begin{center}
\includegraphics[width=0.54\linewidth]{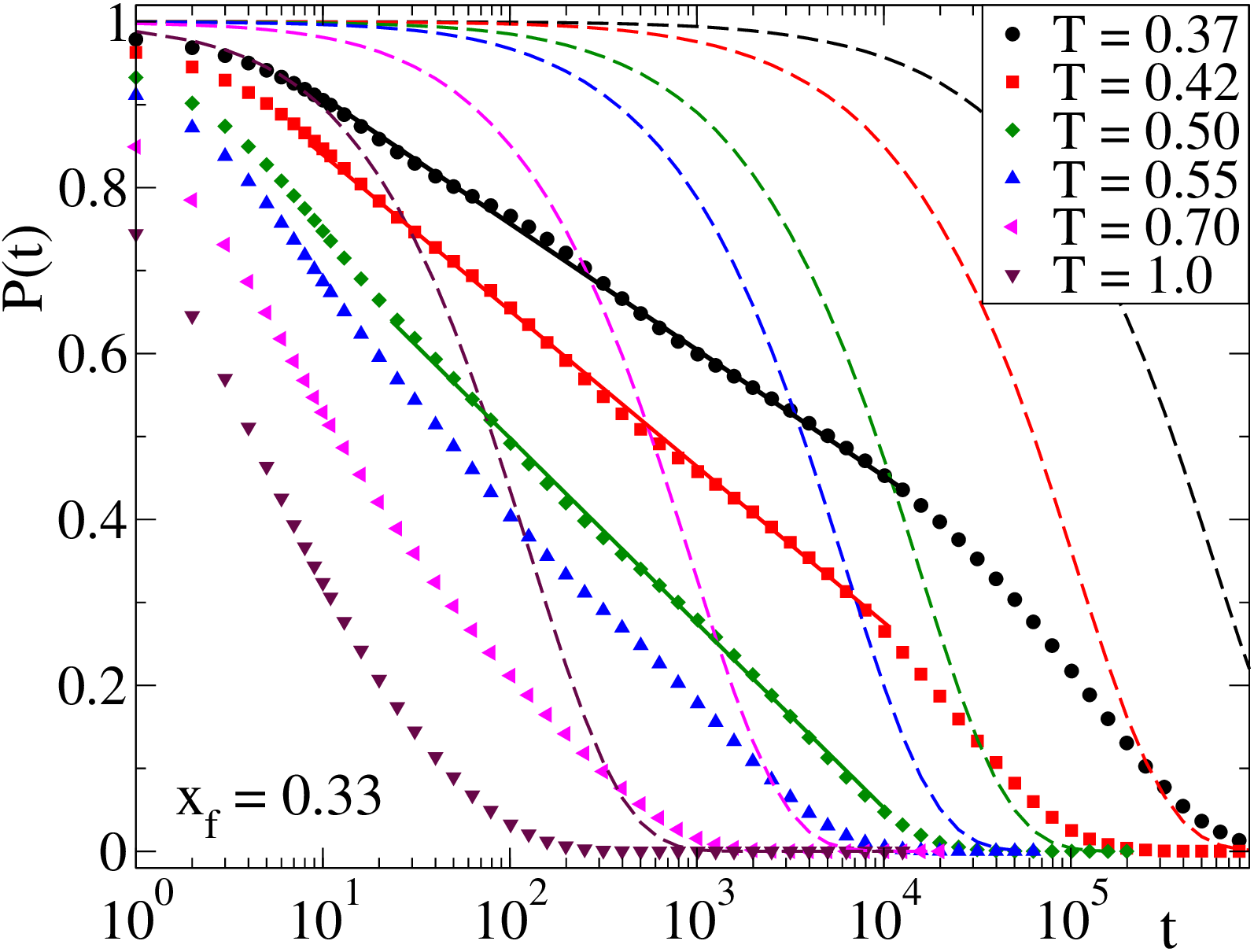}
\caption{Temperature dependence of the mean persistence function $P(t)$ at composition
$x_{\rm f}=0.33$. Symbols correspond to fast cells. Dashed lines correspond to slow cells
(same temperatures as for the former ones, decreasing from left to right).
Solid straight lines indicate logarithmic relaxation.}
\end{center}
\label{fig2}
\end{figure}
\begin{figure}
\begin{center}
\includegraphics[width=0.54\linewidth]{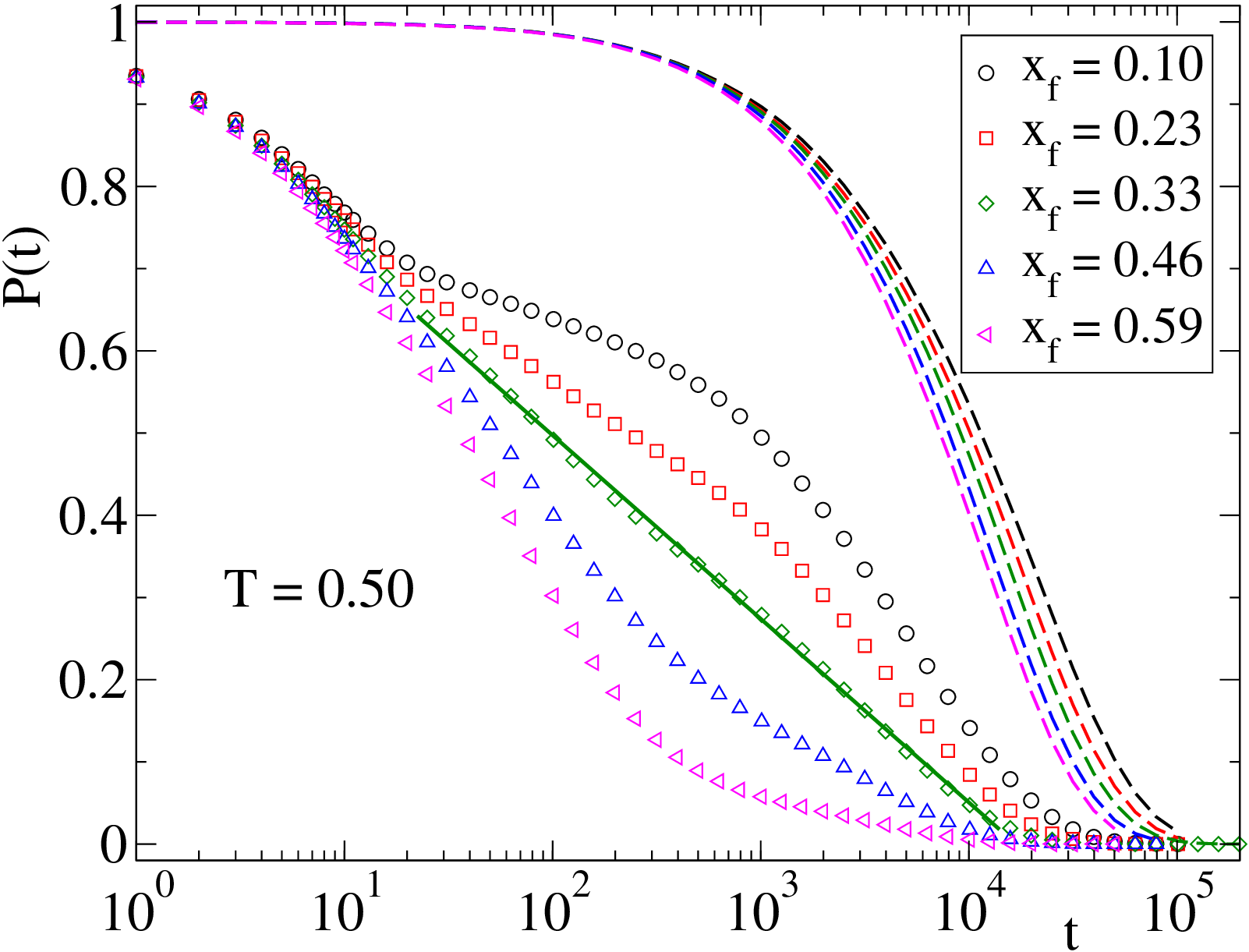}
\caption{Symbols: composition dependence of  $P(t)$, at $T=0.50$, for the fast cells.
Dashed lines correspond to slow cells (same compositions as for the former ones, decreasing from left to right).
The solid straight line indicates logarithmic relaxation.}
\end{center}
\label{fig3}
\end{figure}

It is worthy of remark that the features displayed in Figs. 2 and 3 are not specific of $P(t)$.
Qualitatively analogous results are exhibited by other dynamic correlators, 
as intermediate scattering functions. An example is displayed in Fig. 4 for the incoherent correlator 
$F_{\rm s}(q,t)$. Following the definition given in Ref. \cite{berthier} the latter is computed,
for a wavevector ${\bf q}$, as $F_{\rm s}(q,t) = \langle\exp[i{\bf q}\cdot({\bf r}(t)-{\bf r}(0))]\rangle$, 
where ${\bf r}$ is the position of a probe molecule. The latter is permitted to jump between 
adjacent cells according to the same rules imposed for the evaluation of the mean squared displacement 
of each component (see above). The brackets denote average over initial locations of the probe.
As in the case of the mean persistence function, $F_{\rm s}(q,t)$
exhibits a concave-to-convex crossover by decreasing temperature, and a logarithmic decay
at some intermediate state point.

\begin{figure}
\begin{center}
\includegraphics[width=0.54\linewidth]{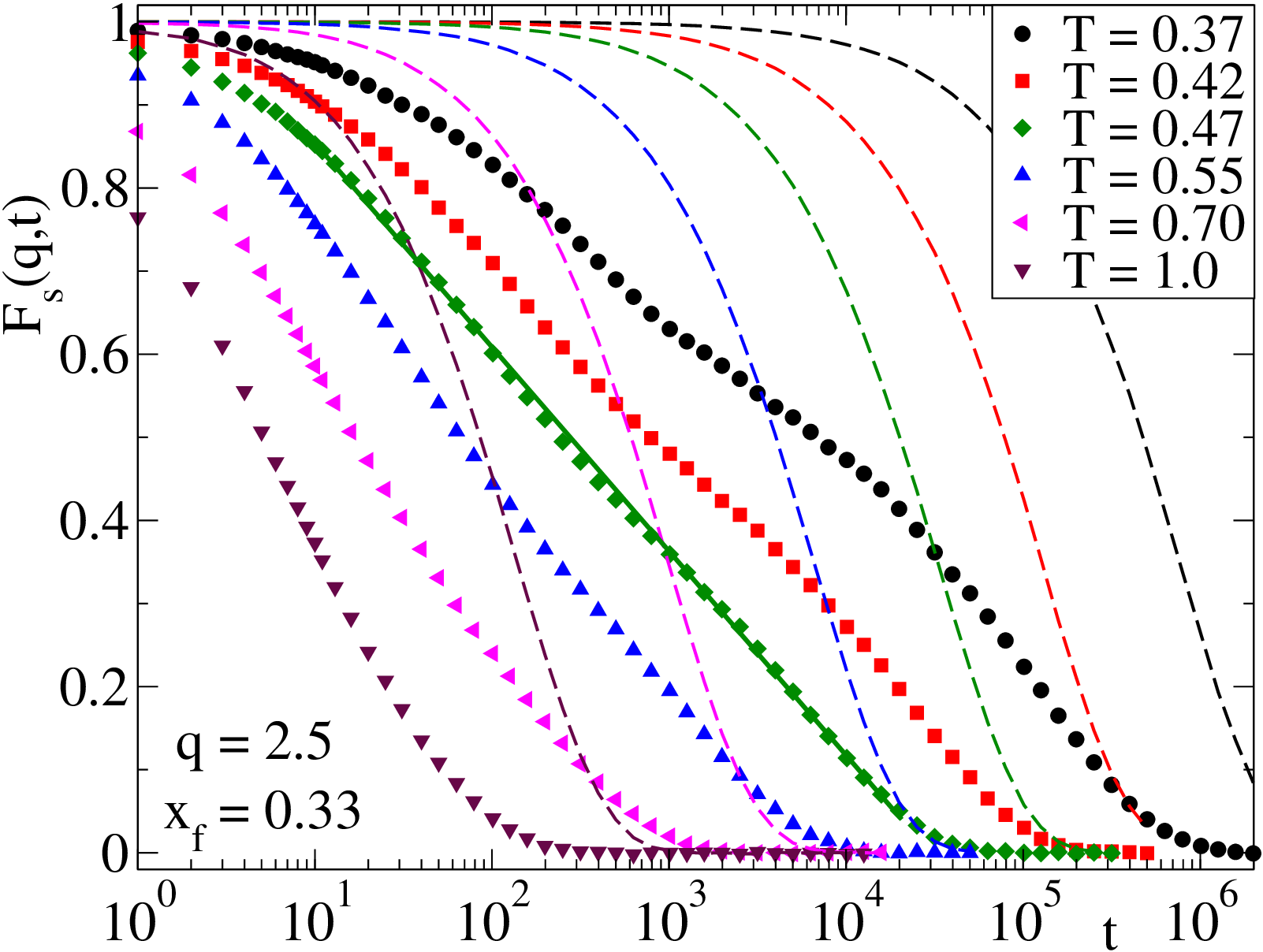}
\caption{Temperature dependence of  $F_{\rm s}(q,t)$, at composition
$x_{\rm f}=0.33$ and wavevector $q = 2.5$. Symbols correspond to fast cells. Dashed lines 
correspond to slow cells (same temperatures as for the former ones, decreasing from left to right)
The solid straight line indicates logarithmic relaxation.}
\end{center}
\label{fig4}
\end{figure}

In summary, the highly non-trivial anomalous relaxation features recently reported for binary mixtures
with a large time scale separation between both components can be qualitatively reproduced 
by a simple kinetically constrained model.
The fact that recent work within this approach \cite{geissler} also reproduces unusual relaxation
features for short-ranged attractive colloids suggests that dynamic facilitation is a relevant
ingredient for relaxation in systems with several competing mechanisms for dynamic arrest.
Results reported here provide a new step to extend the picture of dynamic facilitation,
which had been used basically for describing bulk-like relaxation, to more complex situations.

We thank J. P. Garrahan and L. Berthier for useful discussions.

\end{document}